\documentclass[english,aps,nofootinbib,showpacs,nofootinbib,twocolumn,preprintnumbers]{revtex4}
\usepackage[T1]{fontenc}
\usepackage[latin1]{inputenc}

\usepackage{graphicx}
\usepackage{amsfonts}
\usepackage{amssymb}
\usepackage{calc}
\usepackage{ifthen}
\usepackage{color}

{
{ 
{
\newcommand{\be}{\begin{equation}}
\newcommand{\ee}{\end{equation}}
\newcommand{\bea}{\begin{eqnarray}}
\newcommand{\eea}{\end{eqnarray}}

\newcommand{\bean}{\begin{eqnarray*}}
\newcommand{\eean}{\end{eqnarray*}}

\def\GeV{{\rm \ GeV}}

\def\TeV{{\rm \ TeV}}

\begin{document}
\preprint{ULB-TH/10-08}
\title{Constraining Sommerfeld Enhanced Annihilation Cross-sections of Dark Matter via Direct Searches}
\author{Chiara Arina}
\email{carina@ulb.ac.be}
\author{Fran\c{c}ois-Xavier Josse-Michaux}
\email{fxjossemichaux@gmail.com}
\author{Narendra Sahu}
\email{Narendra.Sahu@ulb.ac.be}
\affiliation{Service de Physique Th\'eorique, Universit\'e Libre de Bruxells, 1050 Brussels,
Belgium}
\begin{abstract}
In a large class of models we show that the light scalar field responsible for the Sommerfeld 
enhancement in the annihilation of dark matter leads to observable direct detection 
rates, due to its mixing with the standard model Higgs. As a result the large annihilation 
cross-section of dark matter at present epoch, required to explain the observed cosmic ray 
anomalies, can be strongly constrained by direct searches. In particular Sommerfeld boost 
factors of order of a few hundred are already out of the CDMS-II upper bound at $90\%$ confidence 
level for reasonable values of the model parameters.

\end{abstract}
\pacs{95.35.+d}
\maketitle

{\bf Introduction - }
Strong evidences support the existence of Dark Matter (DM) in the present Universe~\cite{dm_review}, 
although its actual nature is still missing. Identifying the DM is a major challenge for particle physics 
and has lead to a vast literature on extensions of the Standard Model (SM), in which many 
new particles comply with the requirements that a DM should fulfill. Over the last years, 
many efforts have been dedicated in building models of DM which leave signatures via direct and/or 
indirect detection.

A huge excitement in the indirect detection of DM took place 
after the PAMELA collaboration~\cite{pamela} reported an unexpected rise in the positron fraction at 
energies from 10 GeV up to 100 GeV. Moreover, the H.E.S.S.~\cite{hess} and Fermi Large Area 
Telescope~\cite{fermilat} (FermiLAT) collaborations reported an excess in the electron plus positron flux at 
energies above $100 \GeV$ up to $1 \TeV$.  
If the DM is indeed the source of these observed anomalies in cosmic rays, then for a stable DM the 
current annihilation cross-section should be boosted  
by a factor of $\mathcal{O}(10^{3})$ with respect to the 
freeze-out annihilation cross-section: $\langle \sigma_{\rm DM}|v|\rangle \approx 3 
\times 10^{-26} {\rm cm}^3 {\rm s}^{-1}$. An attractive way of getting a large enhancement of this 
cross-section without affecting the DM relic abundance is to invoke the Sommerfeld effect~\cite{sommerfeld}. 

Recently CDMS-II~\cite{cdms-bound} has reported two events in the signal region at 1.64 $\sigma$ 
confidence level (C.L.). The collaboration has conservatively set an upper bound on the DM spin-independent 
cross-section on nucleon.

\begin{figure}[t!]
\begin{center}
\includegraphics[width=0.45\textwidth]{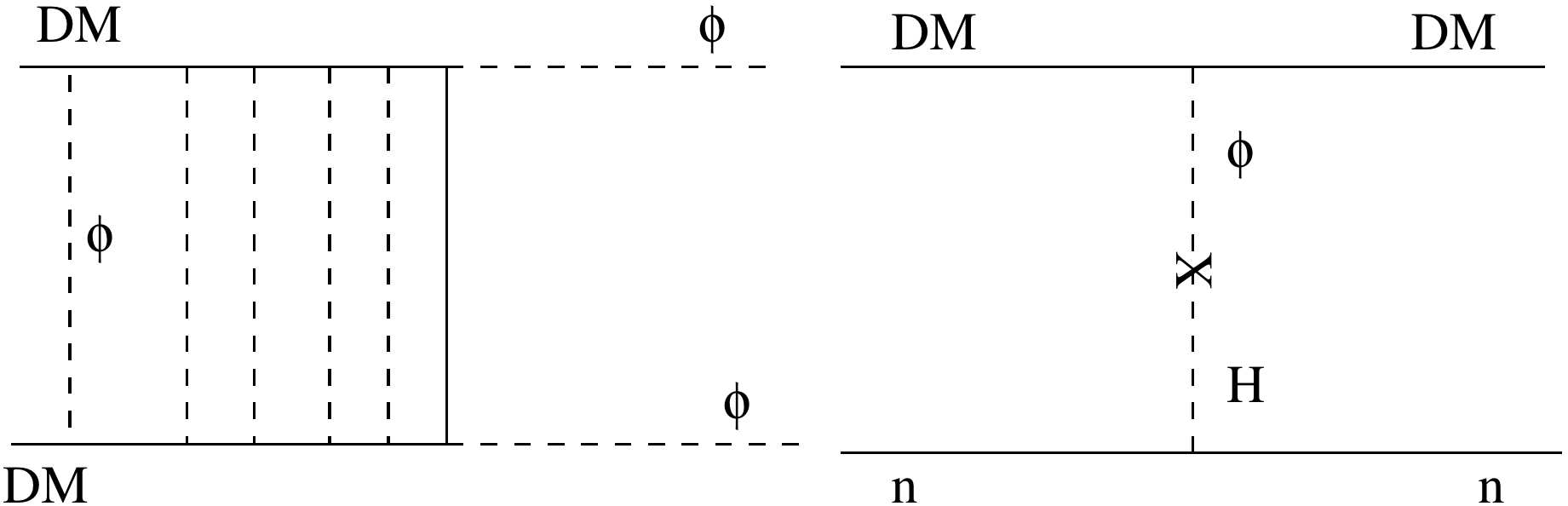}
\caption{A schematic presentation of the role of the light scalar field $\phi$ in indirect (left)
and direct (right) detection of DM.}
\label{fig-1}
\end{center}
\end{figure}

Typically indirect and direct detections of DM probe different sectors of parameter space of a model 
and have been studied as separate issues in the literature. In this letter we consider the 
broad class of model where a light scalar $\phi$ is responsible for the Sommerfeld 
enhancement~\cite{arkani-hamed,west,lightphi,kohri,slatyer}. In these models we show that there 
exists a tight connection between direct and indirect detection of DM. Indeed the coupling responsible 
for the Sommerfeld enhancement also appears in the spin-independent cross-section on nucleon, in a unique 
combination with the Higgs portal coupling ($\phi-H$ mixing, $H$ being the Standard Model Higgs) as 
shown in Fig.~(\ref{fig-1}), Refs.~\cite{slatyer,Chen:2009ab,Hambye:2008bq,hsdm}. We demonstrate
that from the direct detection bounds it is possible to set limits on the Sommerfeld enhancement 
that can arise at present epoch. Alternatively, for a given Sommerfeld enhancement, the Higgs portal 
coupling is strongly constrained~\cite{Chen:2009ab,Cao:2009uv} and can be probed at next generation 
direct detection experiments. 

{\bf Light Scalar - Higgs mixing - } 
Considering that the DM candidate $\chi$ is a SM singlet fermion with mass $M_{\chi}$, stabilized by a 
$Z_2$ symmetry, and that the field $\phi$ is a real singlet scalar, the relevant terms in the Lagrangian 
are:
\bea
-\mathcal{L}&\supset& \lambda_{\chi}\, \overline{\chi^{c}}\chi\phi+\mu_{\phi} \phi H^{\dagger}H\,.
\label{Lagrangian}
\eea
Once $H$ develops a non-zero vacuum expectation value (vev) $v$, $H$ and $\phi$ mix 
through the trilinear term $\propto \mu_{\phi} v$, $v=246\GeV$. Due to $H-\phi$ mixing, 
the scalar field $\phi$ is unstable and ultimately decays to SM fields. Since $\phi$ is produced 
in the current DM annihilation, we demand $m_\phi \lesssim 1\GeV$ in order to avoid the antiproton 
problem. Moreover, in order not to spoil the Big Bang Nucleosynthesis (BBN) predictions the thermally 
generated $\phi$ particles should disappear before the onset of the BBN, thus requiring 
$m_{\phi} > 10$ MeV~\cite{BBN,Hisano:2009rc}. It is true that for a DM candidate with mass in the $100\GeV-1\TeV$ 
range, such a low mass scale of $\phi$ may appear somewhat unnatural, being up-lifted by radiative 
corrections. A supersymmetric realization could however naturally solve this problem. In this letter, 
without addressing this issue, we assume $m_\phi$ to be stabilized at the ${\mathcal O}(1) (\GeV)$ scale.

Considering the Lagrangian in Eq.~(\ref{Lagrangian}), the connection between the direct 
detection and the Sommerfeld enhancement is schematically presented in Fig.~(\ref{fig-1}), with a key 
role played by $\phi$. The trilinear coupling $\lambda_{\chi}\, \overline{\chi^{c}}\chi\phi$ gives rise 
to an attractive Yukawa potential between $\chi$ particles, thus enhancing the current annihilation of 
$\chi \overline{\chi^{c}}\to \phi \phi$. The same coupling is also responsible for a spin independent 
interaction of $\chi$ with the nucleon $n$ through the Higgs portal coupling $\mu_\phi$. We express this 
coupling in terms of the mixing angle between $H$ and $\phi$, $\theta_{H\phi} \sim \mu_{\phi} v/(m^2_H)$, 
where $m_H$ is the physical Higgs mass in the SM. The mixing angle is lower bounded $\theta_{H\phi} 
\gtrsim 10^{-7}/\sqrt{m_{\phi}/\GeV}$ by demanding that the lifetime of $\phi$ should be less then 
$\tau_{\rm BBN} \sim 1$ s~\cite{BBN,Hisano:2009rc}.  The scalar field $\phi$ is indeed thermally produced in the early Universe and should decay before the onset of the primordial nucleosynthesis to avoid dominating the energy density of the Universe~\cite{Chen:2009ab}.
For the mass range of $\phi$ we consider, the mixing between $\phi$ 
and $H$ is also upper bounded, $\theta_{H\phi} < 10^{-2}$~\cite{O'Connell:2006wi}.

The terms in Eq.~(\ref{Lagrangian}) are ubiquitous in hidden sector models. The same Lagrangian arises in 
the case of $\phi$ being a complex singlet and developing a vev. The vector DM and the scalar DM 
cases are very similar to the fermionic example, although the presence of extra couplings may modify the 
phenomenology of the model. If a dark sector is gauged under a hidden symmetry, either abelian or non-abelian, 
then the DM may be constituted by the additional vector gauge boson of the theory~\cite{Hambye:2008bq,Gu:2009hu}. 
Typically, SM fields are singlets under the hidden symmetry. For both abelian and non-abelian cases, the extra 
gauge bosons will couple to the light scalar $\phi$ through the kinetic term 
$\vert (\partial_{\mu}-i g_H T^a A^{'a}_{\mu})\phi\vert^{2}$, $\phi$ being non-singlet under the hidden gauge 
group. The fermionic DM case then just translates to the gauge boson one by replacing $\lambda_{\chi}$ with 
the hidden sector gauge coupling $g_{H}$. In the non-abelian case a kinetic mixing term between 
the SM hypercharge and the hidden gauge sector can arise from higher order operators. In this case interesting 
signatures arise~\cite{Chen:2009ab,hvdm}. In the case the DM is a complex 
scalar field $S$ in the hidden sector, the generality of our argument is somewhat weakened due to the presence 
of the additional coupling $f_{HS} S^{\dagger}SH^{\dagger}H$. The DM directly communicates to the SM sector 
through its Higgs portal coupling. Thus there is an additional channel in the DM to nucleon scattering, 
mediated by the SM Higgs through $t$-channel. Due to the presence of $f_{HS}$, the connection is more 
involved with respect to fermionic and vectorial cases and will be discussed elsewhere~\cite{ourpaper}.

{\bf Sommerfeld enhancement - } 
In the present case, the Sommerfeld enhancement is provided by the light scalar field $\phi$, which 
acts as a long range attractive force carrier between the $\chi$ particles. For a review on the Sommerfeld 
effect due to a light scalar field, see Ref.~\cite{arkani-hamed}, while for non-abelian massive 
vector fields one can see Ref.~\cite{cirelli2007}. We define the coupling constant between 
the DM and $\phi$ as $\alpha_\chi=\lambda_\chi^2/(4 \pi)$. When the Compton wavelength 
${\cal O}(m_\phi^{-1}$) associated with $\phi$ becomes larger than $(\alpha_\chi M_\chi)^{-1}$, the 
asymptotic plane wave $\psi$ associated with $\chi$ gets distorted. The distortion can be computed 
by solving the Schr\"odinger equation for the attractive Yukawa potential $V(r)=-(\alpha_\chi/r)\,
e^{-m_\phi\,r}$ and is defined as $S_{e}=\vert\psi(\infty)/\psi(0)\vert^{2}$~\cite{cirelli2007,arkani-hamed}. 
This is equivalent, in terms of Feynman diagrams, to the resummation of the multi-loop scalar ladder 
contributions, as shown in Fig.~(\ref{fig-1}), the left diagram. The boost factor $S_e$ is a 
function of the dimensionless parameters $\epsilon_{\phi}=m_\phi/(M_{\chi} \alpha_\chi)$ and 
$\epsilon_v = \beta/\alpha_\chi$ only, with $\beta= v_{\rm rel}/c$, the relative velocity between the 
DM particles.
\begin{figure}[t!]
\begin{center}
\includegraphics[width=0.45\textwidth]{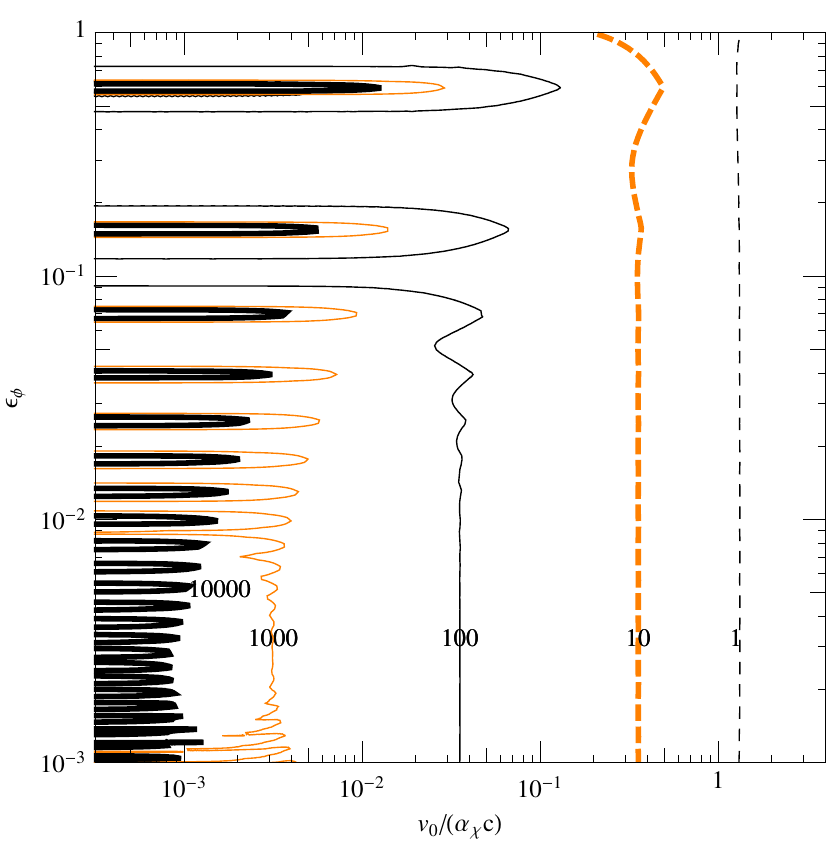}
\caption{Iso-contours of the non-perturbative Sommerfeld enhancement, $\langle S_e \rangle$, as a 
function of $v_0/(\alpha_\chi c) $ and $\epsilon_{\phi}$.}
\label{fig:somm}
\end{center}
\end{figure}

By considering $S_e$ in a halo, we need to integrate over the relative velocity distribution of the 
DM halo~\cite{arkani-hamed,Bovy:2009zs}. Assuming an isothermal Maxwellian distribution with a 
mean velocity $v_0$, the averaged value of $S_e$ reads:
\begin{equation}
\langle S_e \rangle = \frac{4}{\sqrt{\pi}} \Big(\frac{\alpha_\chi c}{v_0}\Big)^3 \int_0^\infty d 
\epsilon_v \,\epsilon^2_v \, {\rm e}^{(- \epsilon^2_v \alpha_\chi^2 c^2/v_0^2)} 
S_e(\epsilon_v, \epsilon_\phi)\,.
\end{equation}
It is a function of $\epsilon_\phi$ and $v_0/(c \alpha_\chi)$ only. We neglect the truncation of the 
Maxwellian distribution, since the cut-off on the escape velocity does not significantly affect 
the results. Indeed the enhancement drops rapidly with increasing velocities. The corresponding iso-contours of 
$\langle S_e \rangle$ are shown in Fig.~(\ref{fig:somm}). For small values of $v_0/(\alpha_\chi c)$ the 
boost factor can rise up to $10^4$, while for large $v_0/(\alpha_\chi c)$ the boost decreases down to 1.

Galactic cosmic rays, being measured by indirect DM search experiments, could arise from the current 
annihilation of DM. As a result PAMELA and Fermi experiments give upper bounds on the present DM total 
annihilation cross-section from the measurement of $e^{+/-}$ and $p-\overline{p}$ fluxes. For a DM mass 
ranging from $100\GeV$ to $1\TeV$, the boost factor can be allowed up to a factor $\mathcal{O}(1000)$. 
Interestingly, as we see below, these boost factors suffer stringent constraints from direct detection 
exclusion limits.

{\bf Direct detection - }
The light scalar $\phi$ is the only field responsible for the DM scattering on nucleon due to its 
mixing with the SM Higgs. The elastic cross-section on nucleon, mediated by $\phi$ through 
$t$-channel, is given by:
\begin{equation}
\sigma_n^{SI} = \frac{\mu^2_n f_n^2 m_n^2}{4 \pi v^2}\frac{\lambda_{\chi}^2 \theta_{H\phi}^2}{m^4_{\phi}}\,,
\label{eq:dd_si}
\end{equation}
where $\mu_n$ is the reduced nucleon-DM mass, $f_n=0.3$, the effective Higgs nucleon interaction. The 
behavior of $\sigma_n^{SI}$ is driven by $1/m^4_\phi$. For a mass of $\phi$ in the MeV-GeV range, 
this cross-section can exceed the current upper bound on the elastic cross-section given by CDMS-II. 
The lighter the $\phi$ is, the smaller is the mixing angle required to be compatible with direct 
DM searches. 

Actually what is measured in a terrestrial DM detector is the differential rate of nuclear recoils, 
integrated over the energy range of the experiment. This quantity is a function of the inverse mean 
velocity of the DM particles that can deposit a given recoil energy $E_r$. For details about 
the total rate the reader is referred to~\cite{totrate} and the references therein. With respect to the 
parameters used to define $\langle S_e \rangle$, the differential rate can be rewritten as:
\begin{equation}
\frac{d R}{dE_r} \propto  \frac{\mu^2_n f_n^2 m_n^2}{v^2} \frac{\lambda_{\chi}^2 c}{4 \pi v_0} 
\frac{\theta_{H\phi}^2}{m^4_{\phi}}\,\frac{1}{y} F(x,y,z)\,.
\label{eq:totrat}
\end{equation}
The function $F$ depends on the minimum velocity to produce a given recoil energy 
$x=v_{\rm min}/v_0$, the observed velocity $y=v_{\rm obs}/v_0$ and the escape velocity 
$z=v_{\rm esc}/v_0$. The observed velocity takes into account the motion of the Earth around the 
Sun. Typically, for an isothermal Maxwellian halo the range of values are $170\, {\rm km/s} < v_0 < 
270\, {\rm km/s}$~\cite{v0} and $498\, {\rm km/s}< v_{\rm esc} < 608\, {\rm km/s}$~\cite{vesc}. The total rate 
is sensitive to the values of the escape velocity, the mean velocity and the minimum velocity, which 
depends on the DM mass, $E_r$ and the nucleus mass. 

We consider only the CDMS-II exclusion limit since it is the most sensitive one in the DM range we 
are interested in (10 GeV-1 TeV). The Xenon10~\cite{xenon-bound} is most sensitive in the range 7-20 GeV.
The experimental upper bound is obtained with the maximum gap method~\cite{mgm} and 
is given at $90\%$ C.L..

{\bf Results and discussions -}
\begin{figure}[t!]
\begin{center}
\includegraphics[width=0.45\textwidth]{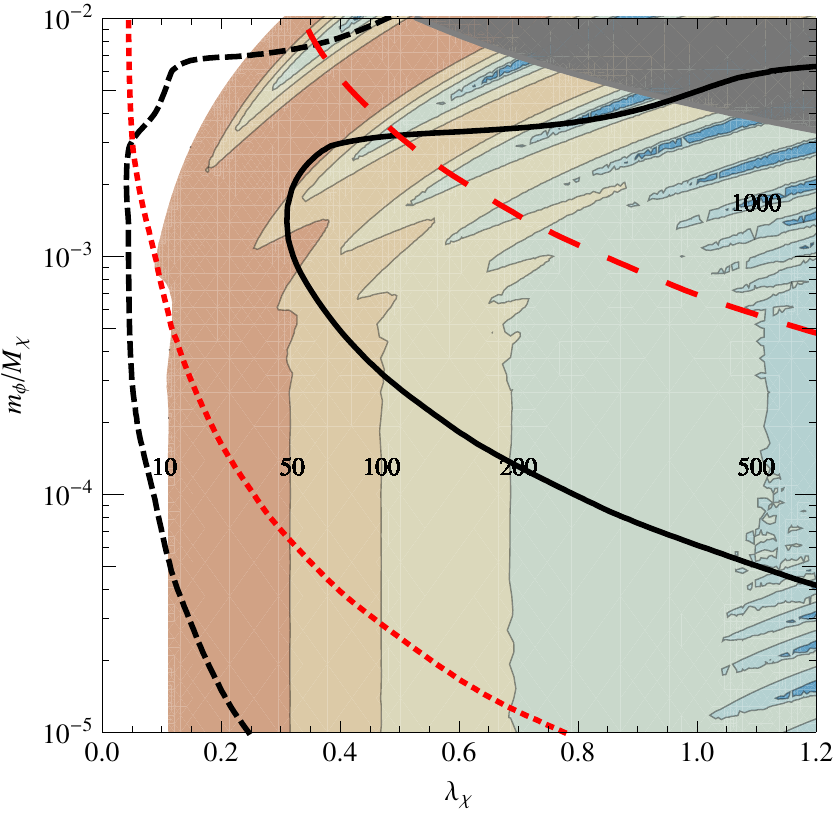}
\caption{Iso-contour of the Sommerfeld enhancement as a function of the coupling $\lambda_{\chi}$ and the 
ratio of the masses $m_\phi/M_{\chi}$. The thick black solid and dashed curves are the exclusion 
limits from CDMS-II for a mixing angle $\theta_{H\phi}=10^{-6}$ and $\theta_{H\phi}=10^{-5}$ respectively, 
with $m_{\phi}= 0.1$ GeV. In analogy the thick red long dashed and dotted lines are for $m_\phi=1$ GeV and 
$\theta_{H\phi}=10^{-4}$ and $\theta_{H\phi}=10^{-3}$ respectively. The right-hand side of each curve is excluded 
at $90\%$ C.L. The gray region is excluded by BBN, Eq.~\ref{eq:BBN}, for $m_\phi=0.25$ GeV.}
\label{fig:wholeb}
\end{center}
\end{figure}
We begin by presenting our results for a mean velocity $v_0$ of the DM particles in the Earth's neighborhood 
of $220$ km/s and an escape velocity of $600$ km/s. The $\langle S_e \rangle$ is therefore a function of 
the coupling $\lambda_\chi$ and $m_\phi/M_{\chi}$. In each graph two suitable values of the mixing angle 
$\theta_{H\phi}$ are taken into account, depending on the $\phi$ mass.

In Fig.~(\ref{fig:wholeb}) we draw the iso-contours of the Sommerfeld boost factor 
and the exclusion limits from CDMS-II for $m_{\phi}= 0.1 \GeV$ (black curves) and 
$1$ GeV (red curves), as labeled in the caption. For $0.1 \leq \lambda_{\chi}\leq 1.2$, boost factors 
ranging from 1 up to 1000 can be obtained, the right range to account for the PAMELA and Fermi $e^{+/-}$ 
and $p - \bar{p}$ measurements. However these values of $\lambda_\chi$ are strongly constrained by direct 
detection limits on DM-nucleon cross-section. For example, if $m_\phi=0.1$ GeV, $M_\chi=100 \GeV$ 
and $\theta_{H\phi}=10^{-6}$ then only small values of $\lambda_\chi$ are allowed, namely $\lambda_\chi 
\lesssim 0.3$. Conversely, the maximum mixing angle allowed from direct detection can be inferred, down 
to $10^{-5}(10^{-3})$ for $m_{\phi}=0.1\GeV (1\GeV)$. 

\begin{figure}[t!]
\begin{center}
\includegraphics[width=0.45\textwidth]{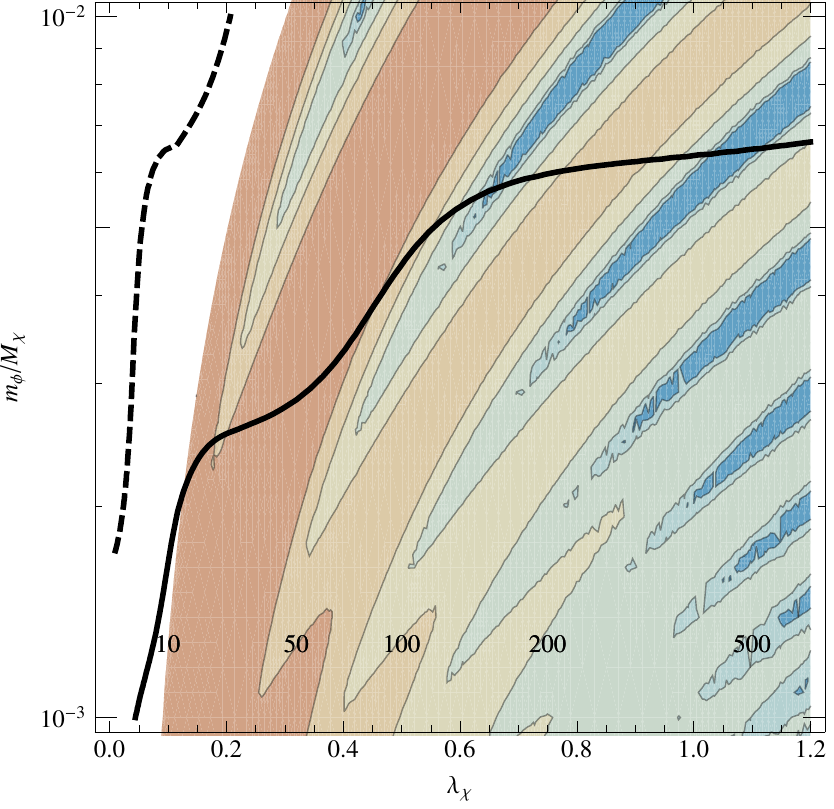}
\caption{Iso-contours of the Sommerfeld enhancement as a function of the coupling $\lambda_{\chi}$ and the 
ratio of the masses $m_\phi/M_\chi$ for a fixed DM mass $M_{\chi}=100$ GeV. The thick solid and 
dashed black curves are the exclusion limits from CDMS-II for a mixing angle $\theta_{H\phi}=10^{-6}$ and 
$\theta_{H\phi}=10^{-4}$ respectively. The region below the curves is excluded at $90\%$ C.L.}
\label{fig:100}
\end{center}
\end{figure}

In Fig.~(\ref{fig:100}) the iso-contours of the Sommerfeld enhancement are depicted and overlapped with 
the bounds from CDMS-II, for a fixed DM mass of $100$ GeV. We can infer the allowed boost factor 
as follows. We see that $\theta_{H\phi} =10^{-6}$ still allows $\langle S_e \rangle > 500$ for value of 
$m_\phi > 0.5$ GeV. A mixing angle of $10^{-4}$ excludes the possibility of having boost factor larger 
than 10. Had we increased the DM mass towards 1 TeV the direct searches would have been less sensitive. 
This is due to the huge difference in mass between the DM particle and the nucleus. For example, the 
CDMS-II detector is made of Ge ($m_{\rm Ge} ~\sim 73 \GeV$), hence the most sensitive region is 
around 50-100 GeV in mass. 

\begin{figure}[t!]
\begin{center}
\includegraphics[width=0.45\textwidth]{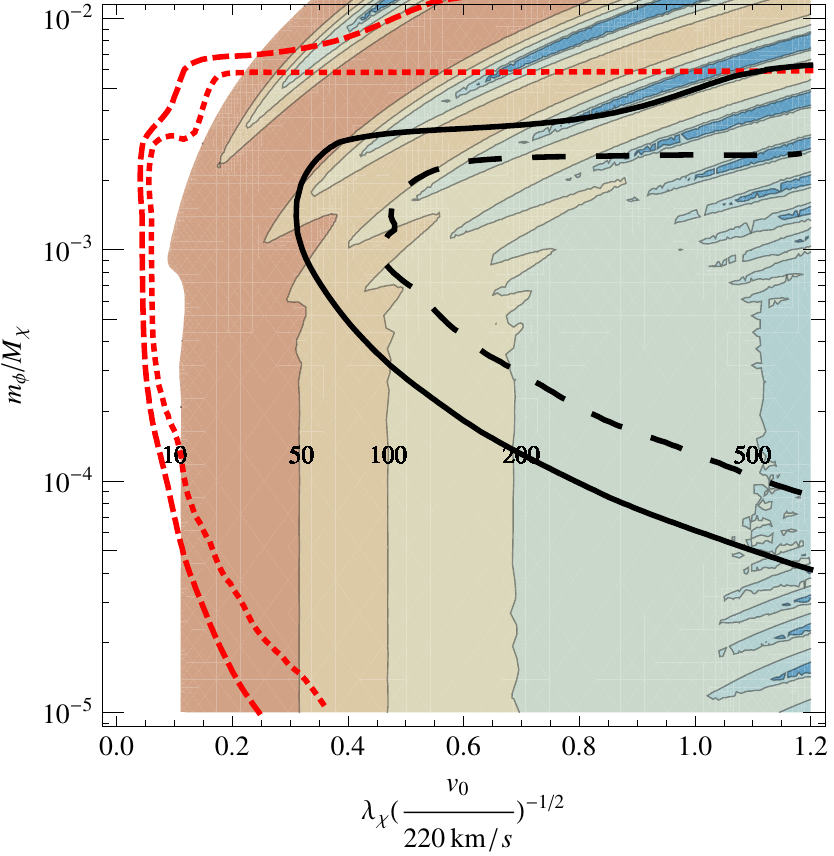}
\caption{Iso-contours of $\langle S_e\rangle$ as a function of the coupling $\lambda_{\chi}/\sqrt{v_0/220\, 
{\rm km/s}}$ and the ratio of the masses $m_\phi/M_{\chi}$ for a fixed $\phi$ mass of 0.1 GeV. The black solid 
and long dashed lines are for $\theta_{H\phi}=10^{-6}$ and two different velocity parameters: $v_0=220$ km/s, 
$v_{\rm esc}=600$ km/s and $v_0=170$ km/s, $v_{\rm esc}=500$ km/s respectively. The same for the dashed 
and dotted red curves with $\theta_{H\phi}=10^{-5}$. The right-hand side of the curves is excluded at $90\%$ C.L..}
\label{fig:170450}
\end{center}
\end{figure}

The effect of the velocity on the exclusion limits and on the Sommerfeld enhancement is better understood 
in terms of the rescaled coupling $\lambda_\chi \sqrt{c/v_0}$. As shown in Fig.~(\ref{fig:somm}), 
$\langle S_e \rangle$ is indeed only dependent on the rescaled coupling. Variations of $v_0$ and 
$v_{\rm esc}$ only affect the direct detection rate through the $F(x,y,z)/y$ factor, see 
Eq.~(\ref{eq:totrat}). This modification is plotted in Fig.~(\ref{fig:170450}) for the extremal 
parameters $v_0=170$ km/s and $v_{\rm esc} = 500$ km/s (red dotted and black long dashed lines) with 
respect to the reference values $v_0=220$ km/s and $v_{\rm esc}= 600$ km/s (red dashed and black solid 
curves). This time we fixed $m_\phi= 0.1$ GeV. The change in direct detection sensitivity may be 
interpreted as a different value of the mixing angle saturating the current experimental upper 
bounds. Larger values of the mixing angle are indeed permitted. For $M_\chi=10$ GeV and 
$\theta_{H\phi}=10^{-5}$, the whole range of $\lambda_\chi$ (red dotted curve) is allowed for the 
extremal velocity parameters, while being upper bounded in the standard case (red dashed line). For 
$\theta_{H\phi}=10^{-6}$ (black curves) and $M_\chi=1000$ GeV, $\lambda_\chi=1.2$ is still permitted 
for the extremal velocity values, implying viable boost factors up to 500 (long dashed), while from the 
reference case (solid) the allowed boost is at most of 200. In other words for a given $\theta_{H\phi}$ 
the maximum allowed $\lambda_\chi$ and boost factor can be inferred. Although the constraints from 
direct detection are slightly reduced for smaller $v_0$ and $v_{\rm esc}$, the main results still hold. 

Notice that we have implicitly assumed an isothermal Maxwellian velocity distribution for describing 
both the Earth's neighborhood (direct detection) and the galactic DM halo ($\langle S_e \rangle$). Our 
results however hold even in the presence of clumpy structures in the DM density profile. In that case, 
from Fig.~(\ref{fig:170450}), we see that the Sommerfeld boost factor is shifted by an amount 
$\sqrt{v_0^{\rm iso}}/\sqrt{v_0^{\rm clump}}$, where $v_0^{\rm clump} \sim 14$ km/s. As a result higher 
boost factors become compatible with the direct detection. 

When the DM interacts with the nucleus at a typical mean velocity $v_0/c\sim 10^{-3}$, a multi-loop 
scalar diagrams, as in Fig.~(\ref{fig-1}), is also present. Potentially it can boost the direct detection 
rate as it does for the annihilation cross-section. With respect to the case of DM annihilation the bound 
state forms between the $\chi$ particle and the nucleus, when they have comparable masses~\cite{Slatyer:2009vg}. 
The attractive Yukawa potential of the light scalar field is proportional to $\lambda_\chi \theta_{H\phi} 
f_n m_n/(4 \pi v)$, due to the mixing with the Higgs boson. The effective coupling giving rise to the 
enhancement is suppressed with respect to $\alpha_\chi$ by the small Higgs nucleon coupling and 
the mixing angle is $\alpha_{DD} \sim 10^{-8}$ for $\theta_{H\phi}\sim 10^{-3}$. In term of 
$\langle S_e \rangle$, from Fig.~(\ref{fig:somm}), this correspond to large values of $v_0/(\alpha_\chi c)$. 
Thus no sizable boost factors are expected in this case. 

We conclude this section with a remark on the relic density of the DM candidate $\chi$ and its link with 
the Sommerfeld boost factor~\cite{fengetal} by assuming that the only interactions are given in 
Eq.~(\ref{Lagrangian}). Then the total annihilation cross-section of $\chi$ can be estimated as: 
\begin{equation}
\langle \sigma_{\chi}|v_{\rm rel}|\rangle \approx \frac{\pi \alpha_\chi^2}{M_\chi^2} \times \Big(1+ \frac{\theta^2_{H\phi}}{2}\Big)
+ \frac{\alpha_\chi \mu_\phi^2}{M_\chi^4} \,.
\label{sigmav}
\end{equation}
Since the mixing angle $\theta_{H\phi} \propto \mu_\phi$, the second and third terms, which give the annihilation 
cross-section of $\chi$ particles into $H\phi$ and $H^\dagger H$, are suppressed for small mixing angles. As a 
result the first term dominates, which gives the annihilation cross-section of $\chi \chi \to \phi \phi$ through 
the $\chi$ exchange in $t-$channel. Hence from the relic abundance of DM, $\Omega_{\rm DM} h^{2} 
\sim 0.1$, one infers $\lambda_\chi \sim 0.6$ for $M_\chi \sim 1 \TeV$, which is in the expected range 
of values that give large Sommerfeld enhancement as well as detectable elastic cross-section on nucleon. 
However, this conclusion does not hold if the $\chi$ particles dominantly annihilate via some other channels 
in the early universe.  

Even after the freeze-out of the DM, a small amount of $\chi$ pairs continue to annihilate with enhanced rate because of the Sommerfeld effect. The increased cross-section may disrupt the $^4\rm He$ and $\rm D$ abundances, which are benchmark predictions of the BBN. In our case the dominant diagram is the $t$-channel $\phi$ production, first term in Eq.~\ref{sigmav}: $\chi\chi \rightarrow \phi\phi$. The $\phi$ subsequently decays mainly into muons, due to its mass range. Assuming that ${\rm BR}(\phi \rightarrow \mu^+\mu^-)$ is $100\%$, the photodissociation of the Helium and Deuterium abundances leads to an upper bound on the Sommerfeld coupling $\lambda_\chi$~\cite{kohri,Hisano:2009rc}:
\begin{equation}
\lambda_\chi \lesssim  0.05 \times\Big(\frac{M_\chi}{\rm GeV}\Big)^{3/4} \Big(\frac{E_{\rm vis}/M_{\chi}}{0.7}\Big)^{-1/4}\,,
\label{eq:BBN}
\end{equation}
where $E_{\rm vis}$ is the energy transferred to the visible sector. Notice that this constraint neither depends on the mixing angle nor on the $\phi$ mass. For a DM mass of 25 GeV, the maximum allowed $\lambda_\chi$ is 0.5, while for $M_{\chi}=100$ GeV $\lambda_\chi = 2.5$. The BBN is capable to set a lower bound on the Higgs portal coupling and moreover for lower DM masses to strongly constraint the Sommerfeld enhancement. In Fig.~\ref{fig:wholeb} the gray region denotes the parameter space excluded by BBN, assuming $m_\phi=0.25$ GeV. For larger DM masses, starting from 50 GeV, the CDMS-II experiment sets the stringent bounds on the allowed boost factors, while BBN constraints are relaxed in the range of values of $\lambda_\chi$ we consider.

At present epoch DM is annihilating in the galactic halo producing charged leptons, with an enhanced rate with respect to the freeze-out rate, due to the Sommerfeld mechanism. An increased neutrino flux can be detected at neutrino telescopes, which will constraint the boost factor~\cite{nu} together with direct searches.

{\bf Conclusions - }
In this letter we analyzed the connection between direct and indirect DM searches in a class of model where a 
light scalar field is added to the SM through a Higgs portal. This scalar acts as a long range force 
carrier and yields the Sommerfeld enhancement for the current annihilation of DM, as required by observed 
cosmic ray anomalies. The crucial point is that through $\phi-H$ mixing the light scalar allows the DM to 
scatter on nucleon, at rates possibly exceeding the exclusion limit of CDMS-II. In such scenarios the direct 
and indirect detections are fully determined only by $\lambda_\chi$, $\theta_{H\phi}$ and $m_{\phi}/M_{\chi}$.
We found that a large part of the parameter space is strongly constrained in order to reconcile the current 
bounds from PAMELA and CDMS-II at $90\%$ C.L.. For example using the $\phi-H$ mixing of order $10^{-6}$, 
$m_{\phi}=0.1$ GeV and $M_{\chi} =1$ TeV, the CDMS-II upper bound does not allow neither a boost factor of 
more than 200 nor $\lambda_{\chi} > 0.75$. While the interplay between direct and indirect detection is 
straightforward in the fermionic example we focused on, we stress that for all models where a light scalar 
is present and the DM candidate has zero SM hypercharge such an interplay does occur.

{\bf Acknowledgements -}
The authors thank T. Hambye, C. Ringeval and M. Tytgat for useful discussions. This work is supported by the IISN and the Belgian Science Policy (IAP VI-11).

\end{document}